\documentstyle[epsfig]{aipproc}

\begin{document}
\title{Spin effects in diffractive hadron photoproduction}

\author{S.V.~Goloskokov}
\address{Bogoliubov Laboratory of Theoretical  Physics,
 Joint Institute for Nuclear Research,
Dubna 141980, Moscow region, Russia}

\maketitle

\begin{abstract}
We study spin asymmetries in diffractive $Q \bar Q$ and vector
meson production which are sensitive to the spin-dependent part
of the two-gluon-nucleon coupling. It is found that the $A_{ll}$
and $A_{lT}$ asymmetry in diffractive reactions can be used to
study polarized gluon distributions of the proton.
\end{abstract}

\section{Introduction}
Investigation of  the structure of hadrons, is a problem of
considerable interest now. The inclusive reaction can be used to
study ordinary parton distributions. However, it is difficult to
distinguish events with a single outgoing proton or jet in a fixed
target experiment like COMPASS \cite{compass}. In this case, at
small $x$ the diffractive events will contribute together with
nondiffractive one. The measured asymmetry can be written in the
form
\begin{equation}\label{as}
A_{exp}=\frac{\Delta \sigma_{ND}+\Delta \sigma_{D}}{\sigma_{ND}+
\sigma_{D}}=A (1-R) + A_D R,\;\; R=\frac{\sigma_{D}}{\sigma_{ND}+
\sigma_{D}}.
\end{equation}
Here $A=\Delta \sigma_{ND}/\sigma_{ND}$ and $A_D=\Delta
\sigma_{D}/\sigma_{D}$. It can be shown that the ratio $R$ should
increase with $x \to 0$. The integrated over $x$ $R$ ratio has
been found at HERA to be about 20--30\% \cite{diff}. This means
that diffractive events might be  important in extraction of
asymmetry  at small $x$ from experiment. The diffractive hadron
photoproduction can be expressed in terms of skewed parton
distribution (SPD) in the nucleon ${\cal F}_\zeta(x)$
\cite{rad-j}. Investigation of such diffractive reactions should
play a keystone role in future study  ${\cal F}_x(x)$ at small
$x$. In the diffractive charm quark production including $J/\Psi$
reactions, the predominant contribution is determined by the
two-gluon exchange (gluon SPD). Analysis of these reactions should
throw light on the gluon structure of the proton at small $x$
\cite{rysk,brod}.

To study spin effects in the diffractive hadron production, one
must know the structure of the two-gluon coupling with the proton
at small $x$.  The QCD--inspired diquark model generates the
spin-dependent $ggp$ coupling \cite{gol_kr} of  the following
form:
\begin{eqnarray}\label{ver}
V_{pgg}^{\alpha\alpha'}(p,t,x_P,l_\perp)&=& (\gamma^{\alpha}
p^{\alpha'} + \gamma^{\alpha'} p^{\alpha}) B(t,x_P,l_\perp)+4
p^{\alpha} p^{\alpha'}
A(t,x_P,l_\perp)\nonumber\\
&+&\epsilon^{\alpha\beta\delta\rho}p_{\delta}\gamma_{\rho}\gamma_{5}
D(t,x_P,l_\perp).
\end{eqnarray}
The first two terms of the vertex (\ref{ver}) are symmetric over
$\alpha,\alpha'$ indices. The structure $(\gamma^{\alpha}
p^{\alpha'} + \gamma^{\alpha'} p^{\alpha}) B(t)$ in (\ref{ver})
determines the spin-non-flip contribution. The term
$p_{\alpha}p_{\alpha'} A(r)$ leads to the transverse spin-flip in
the  vertex which does not vanish in the $s \to \infty$ limit.
The single spin transverse asymmetry predicted in the models
\cite{gol_kr,gol_mod} is about 10\% for $|t| \sim 3 \mbox{GeV}^2$
which is of the same order of magnitude as has been observed
experimentally \cite{krish-f}. These model approaches give for the
ratio  $\alpha=A/B \leq 0.1 \mbox{GeV}^{-1}$

The asymmetric structure in (\ref{ver}) is proportional to $D
\gamma_{\rho}\gamma_{5}$ and can be associated with $\Delta G$.
It should give a visible contribution to the  double spin
longitudinal asymmetry $A_{ll}$ \cite{bartels}. The value of this
structure is not well known now from our model estimations.

In this report, we shall analyze spin effects caused by the
structures $A$ and $B$. It will be shown here that such effects
will be small in the $A_{ll}$ asymmetry.  The double spin
asymmetry for a longitudinally polarized lepton and a transversely
polarized proton is predicted to be not small and mainly
determined by the $A$ term in (\ref{ver}). Such asymmetry should
be used to study this structure in the $ggp$ coupling.

\section{Diffractive hadron production and SPD}
Let us study the diffractive $J/\Psi$ production at high energies
and fixed momentum transfer. The fractions of the momenta of
proton carried by the Pomeron, $\label{x_P} x_P \sim
(m_J^2+Q^2+|t|)/W^2$ is small at high energies. The $\gamma^\star
\to J/\Psi$ transition amplitude is described by a
nonrelativistic wave function \cite{rysk,diehl}. Gluons from the
Pomeron are coupled with the single and different quarks in the
$c \bar c$ loop. The spin-average and spin dependent cross
sections of the $J/\Psi$ leptoproduction with parallel and
antiparallel longitudinal polarization of a lepton and a proton
are determined by the relation
\begin{equation}\label{ds0}
\frac{d \sigma(\pm)}{dQ^2 dy dt} =\frac{1}{2} \left( d
\sigma(^{\rightarrow} _{\Leftarrow}) \pm  d \sigma(^{\rightarrow}
_{\Rightarrow})\right)=\frac{|T^{\pm}|^2}{32 (2\pi)^3
 Q^2 s^2 y}..
\end{equation}
 For the spin-average  amplitude square we find \cite{golj}
\begin{equation}\label{t+}
|T^{+}|^2=  s^2\, N \,\left( (2-2 y+y^2) m_J^2 + 2(1 -y) Q^2
\right) \left[ |\tilde B+2 m \tilde A|^2+|\tilde A|^2 |t| \right].
\end{equation}
 Here the term proportional to $(2-2 y+y^2) m_J^2$ represents
the contribution of the virtual photon with transverse
polarization. The $2(1 -y) Q^2$ term describes the effect of
longitudinal photons. The $N$ factor in (\ref{t+}) is
normalization, and the $\tilde A$ and $\tilde B$ functions are
expressed through the integral over transverse momentum of the
gluon. The function $\tilde B$ is determined by
\begin{eqnarray}\label{bt}
&&\tilde B=\frac{1}{4 \bar Q^2} \int \frac{d^2l_\perp
(l_\perp^2+\vec l_\perp \vec \Delta) B(l_\perp^2,x_P,...)}
{(l_\perp^2+\lambda^2)((\vec l_\perp+\vec
\Delta)^2+\lambda^2)[l_\perp^2+\vec l_\perp \vec \Delta
+\bar Q^2]} \nonumber\\
&\sim& \frac{1}{4 \bar Q^4}\int^{l_\perp^2<\bar Q^2}_0
\frac{d^2l_\perp (l_\perp^2+\vec l_\perp \vec \Delta) }
{(l_\perp^2+\lambda^2)((\vec l_\perp+\vec \Delta)^2+\lambda^2)}
B(l_\perp^2,x_P)=\frac{1}{4 \bar Q^4} {\cal F}^g_{x_P}(x_P,t,\bar
Q^2).
\end{eqnarray}
and connected with the gluon SPD \cite{kroll-da}. Here $\bar
Q^2=(m_J^2+Q^2+|t|)/4$.  The $\tilde A$ function is determined by
the similar integral.
\begin{figure}[b!] 
\centerline{\epsfig{file=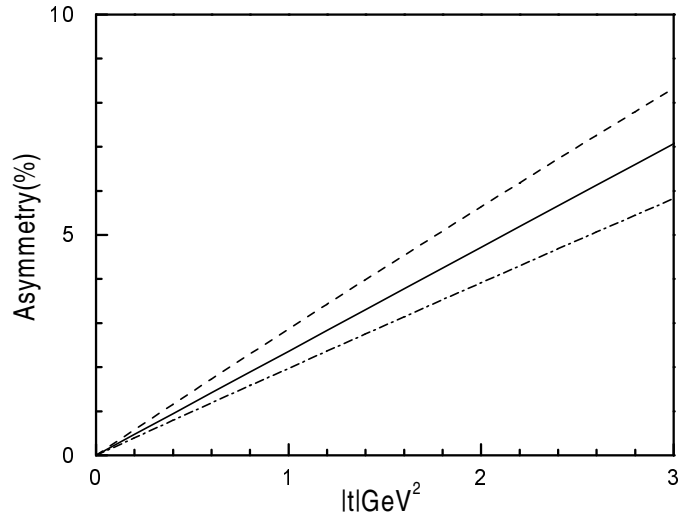,width=3.5in}}
\vspace{10pt} \caption{The $A_{ll}$ asymmetry of the $J/\Psi$
production  at HERMES: solid line -for $\alpha_{flip}=0$;
dot-dashed line -for $\alpha_{flip}=-0.1$; dashed line -for
$\alpha_{flip}=0.1$} \label{fig1}
\end{figure}
The spin-dependent amplitude square looks like
\begin{equation}\label{t-}
|T^{-}|^2= s |t|  (2- y) N   \left[ |\tilde B|^2+ m (\tilde
A^\star \tilde B +\tilde A \tilde B^\star) \right] m_J^2.
\end{equation}

The asymmetry $A_{ll}=\sigma(-)/\sigma(+)$  depends on the ratio
of the spin-flip to the non-flip  parts of the coupling
(\ref{ver}) $\alpha_{flip}=\tilde A(t)/ \tilde B(t)$ which has
been found  to be about 0.1. The predicted asymmetry at HERMES
energies is shown in Fig.\ 1. The contribution of the
spin-dependent $A$ term in (\ref{ver}) to the double spin
$A_{ll}$ asymmetry of the $J/\Psi$ production does not exceed two
per cent for the momentum transfer $|t| \leq 1 \mbox{GeV}^2$.
Sensitivity of the asymmetry to $\alpha$ is rather weak.   At
HERA energies, the asymmetry will be negligible.

For the diffractive $Q \bar Q$ leptoproduction the spin-average
and spin-dependent cross section can be written in the form
\begin{equation}
\label{sigma} \frac{d^5 \sigma(\pm)}{dQ^2 dy dx_P dt dk_\perp^2}=
\left(^{(2-2 y+y^2)} _{\hspace{3mm}(2-y)}\right)
 \frac{C(x_P,Q^2) \; N(\pm)}
{\sqrt{1-4k_\perp^2\beta/Q^2}}.
\end{equation}
Here $C(x_P,Q^2)$ is a normalization function which is common for
the spin average and spin dependent cross section. The $N(\pm)$
functions are expressed through the same skewed gluon
distributions ${\cal F}^g_{x_P}(x_P,t,\bar Q_1^2)$ as for vector
meson production but at a different scale $Q_1^2=m_Q^2+k_\perp^2$
(k is a quark momentum). Note that $x_P$ is not fixed in this
reaction and usually $x_P \leq 0.1$
\begin{figure}[t!] 
\centerline{\epsfig{file=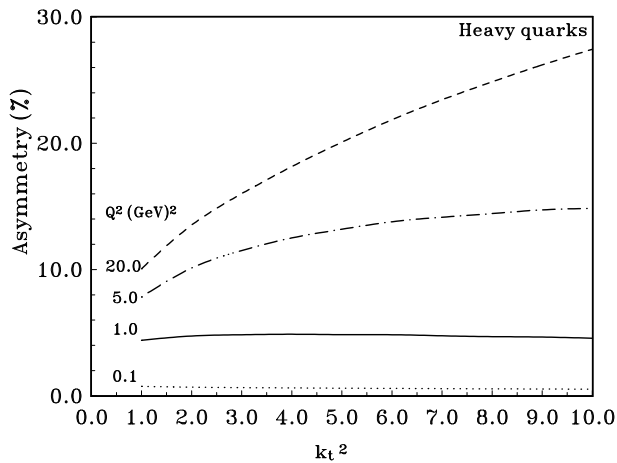,width=3.5in}}
\vspace{10pt} \caption{The predicted $Q^2$ dependence of the
$A_{lT}$ asymmetry for the $c \bar c$ production at COMPASS for
$\alpha=0.1\mbox{GeV}^{-1}$, $x_P$=0.1, $y$=0.5} \label{fig2}
\end{figure}
The predicted asymmetry is quite small and does not exceed 1\%.
It has a week dependence on the $\alpha=\tilde A/\tilde B$ ratio.
Moreover, the $A_{ll}$ asymmetry is predicted to vanish for $Q^2
\to 0$  as $A_{ll} \propto Q^2/(Q^2+Q^2_0)$ with $Q^2_0 \sim 1
\mbox{GeV}^2$.

The $A$ structure in (\ref{ver})  should contribute to the
$A_{lT}$ asymmetry with longitudinal lepton and transverse proton
polarization. The calculation of this asymmetry is similar to the
analysis of $A_{ll}$, which has been carried out before. It has
been found that the $A_{lT}$ asymmetry is not small and
proportional to the $\alpha=\tilde A/\tilde B$ ratio. The
$A_{lT}$ asymmetry is proportional to the scalar production of
the proton spin vector, and the jet momentum $A_{lT} \propto
(s_{\perp} \cdot k_{\perp}) \propto \cos (\phi_{Jet})$ and the
asymmetry integrated over the azimuthal jet angle $\phi_{Jet}$ is
zero. We have calculated the $A_{lT}$ asymmetry for the case when
the proton spin vector is perpendicular to the lepton scattering
plane and the jet momentum is parallel to this spin vector. The
predicted asymmetry is large and shown in Fig. 2. The reason for
the large value of $A_{lT}$ is that we do not find here a small
coefficient $x_P$ as for the $A_{ll}$ asymmetry \cite{gola_ll}.
\section{Conclusion}
In the present report, the polarized cross section of the
diffractive hadron leptoproduction at high energies has been
studied. The two-gluon exchange model with the spin-dependent
$gg$-proton coupling (\ref{ver}) has been used. We consider all
the graphs where the gluons from the Pomeron couple to a
different quark in the loop and to the single one. This provides
a gauge-invariant scattering amplitude.

Our calculations show that the contribution of the structure $A$
in (\ref{ver}) to $A_{ll}$  is smaller than 1-2\%. Not small
effects in the double spin $A_{ll}$ asymmetry should be
determined by the $\Delta G \propto D \gamma_{\rho}\gamma_{5}$
term of the vertex (\ref{ver}). The results obtained here show
that diffractive asymmetry in the $Q \bar Q$ production vanishes
as $Q^2 \to 0$.  We can conclude that most likely such effects do
not provide additional problems in extracting $\Delta G$ from the
$A_{ll}$ asymmetry because  the COMPASS experiment plans to study
the open charm production at small $Q^2$ \cite{compass}.

It is shown that the gluon SPD ${\cal F}^g_{x_P}(x_P)$ and
connected with $\Delta G$ distribution ${\cal G}^g_{x_P}(x_P)$ at
the small $x_P \sim (m_V^2+Q^2)/W^2$ can be studied from the
double spin asymmetry in the vector meson photoproduction. The
contributions of the quark SPDs are non-negligible for $x$ of
about 0.1 where the HERMES and COMPASS experiments will operate.
Thus, in the case of the $\phi$ production the strange quark SPD
might be studied in addition to the gluon one.

It is found here that the $A_{lT}$ asymmetry of the diffractive
heavy quark production is predicted to be not small, about
10-20\%. It can give direct information about the spin-dependent
structure $A$ in the $ggp$ coupling. A similar contribution to
$A_{lT}$ in the vector meson production vanishes because of the
integration over $k_\perp$. The structure, which is proportional
to $x_p$ in the $A_{lT}$ asymmetry of the vector meson production
will be studied later.

We can conclude that important information on the spin--dependent
SPD at small $x$ can be obtained from double spin asymmetries in
diffractive hadron photoproduction reactions .

The author is grateful to the Organizing Committee of SPIN2000 for
financial support. This work was supported in part by the Russian
Fond of Fundamental Research, Grant 00-02-16696.

\end{document}